\documentclass[twocolumn,twoside,slac_two]{revtex4}
\usepackage{graphicx}
\usepackage{fancyhdr}
\usepackage{mathrsfs}
\def \rightdownarrow
 {\kern.3em
 \rule[.5ex]{.15mm}{2ex}
 {\mbox{$\kern-0.1em{\longrightarrow}$}}      
 }

\def\lessim{\mathrel {\vcenter {\baselineskip 0pt \kern 0pt  
\hbox{$<$} \kern 0pt \hbox{$\sim$} }}}

\def\gessim{\mathrel {\vcenter {\baselineskip 0pt \kern 0pt   
\hbox{$>$} \kern 0pt \hbox{$\sim$} }}}

\newcommand{\ie}{i.\ e.,}					
\newcommand{\eg}{e.\ g.,}					

\newcommand{\lumi}{\mbox{cm$^{-2}$s$^{-1}$}}  	 		
\newcommand{\Lumi}{\ensuremath{\mathcal{L}}}			
\newcommand{\lumipb}{\mbox{pb$^{-1}$}}				
\newcommand{\lumifb}{\mbox{fb$^{-1}$}}				

\newcommand{\br}{\ensuremath{\mathcal{B}}}

\newcommand{\tev}{\ensuremath{\mathrm{Te\kern -0.1em V}}}
\newcommand{\gev}{\ensuremath{\mathrm{Ge\kern -0.1em V}}}	
\newcommand{\mev}{\ensuremath{\mathrm{Me\kern -0.1em V}}}	
\newcommand{\kev}{\ensuremath{\mathrm{ke\kern -0.1em V}}}	
\newcommand{\massmev}{\mbox{\mev/$c^2$}}			
\newcommand{\pgev}{\mbox{\gev/$c$}}				

\newcommand{\CP}{\ensuremath{\mathsf{CP}}}			
					
\newcommand{\vub}{\ensuremath{V_{ub}}}

\newcommand{\vtd}{\ensuremath{V_{td}}}
\newcommand{\vts}{\ensuremath{V_{ts}}}

\newcommand{\pap}{\proton\antiproton}			
\newcommand{\pt}{\ensuremath{p_{\rm{T}}}}			
\newcommand{\ptb}{\ensuremath{\pt(B)}}				

\newcommand{\cdfii}{CDF II}

\newcommand{\proton}{\ensuremath{\mathrm{p}}}
\newcommand{\antiproton}{\ensuremath{\bar{\rm{p}}}}

\newcommand{\bab}{\ensuremath{b\bar{b}}}

\newcommand{\Yquattros}{\mbox{$\Upsilon$(4S)}}
\newcommand{\Ycinques}{\mbox{$\Upsilon$(5S)}}

\newcommand{\bd}{\ensuremath{B^{0}}}				
\newcommand{\bs}{\ensuremath{B^{0}_s}}				
\newcommand{\bu}{\ensuremath{B^{+}}}				
\newcommand{\bc}{\ensuremath{B^{+}_c}}				

\newcommand{\abd}{\ensuremath{\overline{B}^{0}}}		

\newcommand{\bquark}{\mbox{$b$-quark}}				

\newcommand{\bhadron}{\mbox{$b$-hadron}}			
\newcommand{\bhadrons}{\mbox{$b$-hadrons}}			

\newcommand{\bbaryons}{\mbox{$b$-baryons}}			

\newcommand{\bgmeson}{\mbox{$b$-meson}}				
\newcommand{\bgmesons}{\mbox{$b$-mesons}}			

\newcommand{\bn}{\ensuremath{B^{0}_{(s)}}}			
\newcommand{\bnmeson}{\mbox{$B^0_{(s)}$ meson}}			

\newcommand{\bdmeson}{\mbox{$B^0$ meson}}
\newcommand{\bsmeson}{\mbox{$B^0_{s}$ meson}}

\newcommand{\bmumu}{\ensuremath{\bn \rightarrow \mu^{+}\mu^{-}}}
\newcommand{\bsmumu}{\ensuremath{\bs \rightarrow \mu^{+}\mu^{-}}}
\newcommand{\bdmumu}{\ensuremath{\bd \rightarrow \mu^{+}\mu^{-}}}

\newcommand{\bhh}{\ensuremath{\bn \rightarrow h^{+}h^{'-}}}

\newcommand{\bdpipi}{\ensuremath{\bd \rightarrow \pi^+ \pi^-}}

\newcommand{\bdkpi}{\ensuremath{\bd \rightarrow K^+ \pi^-}}
\newcommand{\abdkpi}{\ensuremath{\abd \rightarrow K^- \pi^+}}

\newcommand{\bskpi}{\ensuremath{\bs \rightarrow K^- \pi^+}}

\newcommand{\bskk}{\ensuremath{\bs \rightarrow  K^+ K^-}}

\newcommand{\bspipi}{\ensuremath{\bs \rightarrow  \pi^+ \pi^-}}

\newcommand{\jpsi}{\ensuremath{J/\psi}}

\newcommand{\dzero}{\ensuremath{D^{0}}}

\newcommand{\dplus}{\ensuremath{D^{+}}} 		  

\newcommand{\dzerokpi}{\ensuremath{\dzero \rightarrow K^- \pi^+}}

\newcommand{\dpluskpipi}{\ensuremath{\dplus \rightarrow K^- \pi^+ \pi^+}}

\newcommand{\fig}[1]{Fig.\ \ref{fig:#1}}

\newcommand{\eq}[1]{Eq.\ (\ref{eq:#1})}

\newcommand{\refcita}[1]{Ref.\ \cite{#1}}

\newcommand{\cita}[1]{\cite{#1}}

\newcommand{\beq}{\begin{equation}}
\newcommand{\eeq}{\end{equation}}  
\newcommand{\beqn}{\begin{eqnarray}}
\newcommand{\eeqn}{\end{eqnarray}}

\newcommand{\dedx}{\ensuremath{\it{dE/dx}}}
\newcommand{\like}{\ensuremath{\mathscr{L}}}
\newcommand{\gauss}{\ensuremath{\mathscr{G}}}
\newcommand{\pdf}{\ensuremath{\wp}}
\newcommand{\ptot}{\ensuremath{p_{\rm{tot}}}}

\newcommand{\acp}{\ensuremath{A_{\mathsf{CP}}}}
\newcommand{\acpbdkpi}{\ensuremath{\acp(\bdkpi)}}

\begin{document}


\title{CDF Hot Topics}

%

\author{D.~Tonelli (for the CDF Collaboration)}
\affiliation{I.N.F.N. Sezione di Pisa - Ed. C, Polo Fibonacci, Largo B. Pontecorvo, 3 - 56127 Pisa, Italy}

\begin{abstract}
After an introduction on the peculiarities of flavor-physics measurements at a hadron collider, and on the upgraded Collider Detector at Fermilab (\cdfii),
I show recent results on two-body \bd\ and \bs\ decays into charged, pseudo-scalar, charmless mesons or into muons, to illustrate how 
the flavor physics program at CDF is competitive with (in \bd\ decays) and complementary (in \bs\ decays) to $B$-factories. Results shown include 
the new measurement of the \CP-violating asymmetry in \bdkpi\ decays, the first measurement of the time-evolution of \bskk\ decays, and the world best
 limits on the decay rates of rare \bmumu\ modes. 
\end{abstract}

\maketitle

\thispagestyle{fancy}


\section{Introduction}
\label{sec:intro}
Precise measurements in \bd\ and \bu\ meson decays, performed in recent years at the $B$-factories,
provided several substantial improvements in understanding flavor dynamics. 
However, several open questions remain, and a continuous experimental 
effort is still necessary to  complete the picture. In this respect, simultaneous access to decays of strange and non-strange \bgmesons,
an opportunity that is currently unique to the Tevatron\footnote{After a short test-run, the Belle Collaboration is considering the opportunity 
to run at the \Ycinques\ resonance in the near future; in this case, they will have access to large samples of \bs\ meson decays as well.}, might simplify the extraction of 
quark flavor-mixing (CKM) parameters in measurements affected by sizable hadronic uncertainties. 
\par I focus here on the recent CDF results on two-body \bn\ decays into charmless, charged pseudo-scalar mesons 
or into muons. These results show how CDF program is becoming fully competitive with $B$-factories in untagged, time-independent analyses 
of \bd\ meson decays into charged final states, and complementary to $B$-factories in analyses of \bs\ modes.
\par Unless otherwise stated,  \textsf{C}-conjugate 
modes are implied throughout this paper, and branching-fractions (\br) indicate \CP-averages.
In addition, the first uncertainty associated to any number is statistical, and the second one is systematic. 

\section{\cdfii\ at the Tevatron collider}
\label{sec:cdf}
The analyses presented here used data corresponding to time-integrated luminosities of $\int\Lumi dt\simeq 360$--780 \lumipb,
collected by the upgraded Collider Detector (\cdfii) at the Fermilab Tevatron Collider.
\subsection{The Tevatron collider}
\label{ssec:tevatron}
The Tevatron  is a superconducting proton-synchrotron at the final stage of the Fermilab accelerator complex.
In Run II (mid-2001--present), it accelerates 36 bunches of protons against 36 bunches of
  anti-protons producing one crossing every 396 ns at $\sqrt{s} = 1.96$ TeV.
Since the interaction region is about 30 cm long (r.m.s.) along the beam-line,
 a properly designed silicon micro-vertex detector is required to ensure good
 coverage for charged particles.  The transverse beam-width at the collision point is about
 25--30 $\mu$m (r.m.s.), sufficiently small with respect to the typical 
transverse decay-length\footnote{$L_{xy}=\beta_T\gamma c \tau$, with 
 typical Lorentz boost projected onto the plane perpendicular to the beam-line $\beta_T\gamma \simeq$ 0.5--2, and 
typical pseudo-proper decay-length of the $b$-hadron $c\tau \simeq 450$ $\mu$m.} of
 $b$-hadrons, $L_{xy}\simeq 450~\mu$m, to allow separation of secondary from primary vertices. 
The instantaneous luminosity (\Lumi) has been rising steadily during Run II up to the world record peak of 
$\Lumi \simeq 1.82\times 10^{32}$ \lumi, and regularly exceeds 
$\Lumi = 10^{32}$ \lumi; at such luminosities, two \pap\ interactions per bunch-crossing occur on average.
The machine typically delivers data corresponding to $\int\Lumi dt \simeq 20$ \lumipb\ per week, which
are recorded with average data-taking efficiencies in excess of 85\% at CDF. 
As of May 2006, the total luminosity gathered on tape is around  1.4 \lumifb, of which approximately 1 \lumifb\  was recorded
with all crucial sub-detectors for flavor-physics operative.

\subsection{The \cdfii\ detector}
\label{ssec:detector}
The \cdfii\ detector is a 5000 t, multipurpose, solenoidal magnetic-spectrometer
 surrounded by  $4\pi$ calorimeters and muon detectors;
 it is axially and azimuthally symmetric around the interaction point.
Its excellent tracking performance, good muon coverage, and  
particle identification (PID) capabilities allow a broad flavor-physics program.
  We briefly outline the sub-detectors pertinent to the analyses
described here, additional details can be found elsewhere \cite{gen_1,gen_2}.
 \par The \cdfii\ tracking system consists of an inner silicon system 
surrounded by a cylindrical gas-wire drift chamber, both immersed in a 1.4 T solenoidal magnetic field with
 135 cm total lever arm. Six (central region, $|\eta| < 1$) to seven (forward, $1< |\eta| < 2$) double-sided silicon layers, plus
 one single-sided layer, extend radially from 1.6 to 22 cm (28 cm) from the beam line in the central (forward) region, fully covering the
 luminous region.
 The chamber provides 96 (48 axial and 48 stereo) samplings of charged-particle paths 
between 40 and 137 cm from the beam, and within $|\eta|< 1$. The long lever-arm of the tracker provides a superb mass-resolution: with
$\sigma_{\pt}/\pt^2 < 0.15\% (\pgev)^{-1}$, typical observed mass-widths are about 14 \massmev\ for $\jpsi\rightarrow\mu^+\mu^-$ decays, and about 9 
\massmev\ for \dzerokpi\ decays. In addition, silicon measurements close to the beam allow precise reconstruction of decay vertices, 
with typical resolutions of 30 $\mu$m in the transverse plane and 70 $\mu$m along the beam direction.\par  
Four layers of planar drift chambers detect muon candidates with $\pt > 1.4$ \pgev\ in the 
$|\eta|< 0.6$ region, while conical sections of drift tubes extend the coverage to $0.6 < |\eta| < 1.0$ for muon candidates
with $\pt > 2.0$ \pgev. Low-momentum PID is obtained with a scintillator-based Time-of-Flight detector with about 110 ps
 resolution,  that provides $2\sigma$ separation between kaons and pions with $p<1.5$ \pgev.
 The information of specific energy-loss from the drift chamber (\dedx) complements the PID with 1.4$\sigma$ 
 nearly constant $K/\pi$ separation for higher-momentum charged particles ($p>2$ \pgev).

\subsection{Flavor physics at CDF}
\label{ssec:flavor}
Heavy-flavor phenomenology at the Tevatron is different
with respect to the $e^{+}e^{-}$ environment. A large $\pap \rightarrow \bab X$ rate is exploited, which results in a production 
cross-section  of about 30 $\mu$b \cite{gen_1} for \bhadrons\ within detector coverage, compared to 1 nb (7 nb) 
$e^+e^-\rightarrow \bab$  cross-sections at the \Yquattros\ ($Z^0$) resonances.  
Unlike at the $B$-factories,  all species of \bhadrons\  are produced at the Tevatron, including \bs\ and \bc\ mesons, and \bbaryons.
In addition, the dominant source of \bhadrons\ is incoherent, strong production of  \bab\ pairs;
thus measurements that require flavor-tagging can be done 
by reconstructing a single \bhadron\ in the event, while at $B$-factories
the flavor of one \bgmeson\ is determined only after observing the decay of the other one.
\par
 However, the \pap\ collider poses also several challenges. The large \bab\ production cross-section
 is still only about 1/1000$^{th}$ the total inelastic \pap\ cross-section.
 Moreover, high ($\approx$ 30)  track multiplicities per event are observed at the Tevatron, owing to fragmentation of
 the hard-interaction products, to the underlying events (\ie\ hadronized remnants 
of \proton\ and \antiproton) and  to pile-up events (multiple collisions per bunch
 crossing). In addition,  the transverse momentum distribution of \bhadrons\   is a rapidly falling function: 
most  \bhadrons\  have low-\pt\ and  decay into particles which are typically quite soft,
often having $\pt < 1$ \pgev; thus, the need to select low-\pt\ particles  conflicts with the limited bandwidth allowed by the data
 acquisition systems; furthermore, since the longitudinal  component  
of \bhadron\  momenta is frequently large, their decay-products tend to be boosted along the beam line, 
thus escaping the detector  acceptance. If one \bquark\ is within CDF acceptance, the other one 
is within acceptance only $\mathcal{O}(10\%)$ of the time.\par 
The best way to identify $b$-flavor decays in such a challenging environment
is to exploit their  relatively long lifetime, which results in decay
vertices separated by hundreds of microns from
 the \pap\ interaction, for hadrons with typical boosts. 
In this respect, CDF excellent tracking plays a key role for an effective \bhadron\  
reconstruction; equally important is its highly-specialized  and selective trigger, which uses
silicon tracking information online and gathers large and pure samples of \emph{charmed} and \emph{beauty} decays while
sustaining the high-rates associated with the \pap\ environment. 

\subsection{Role of trigger}
\label{ssec:trigger}
CDF exploits its unique ability to trigger events
with charged particles  originated in vertices displaced from the
primary \pap\ vertex (displaced tracks) \cite{gen_3}. 
Displaced tracks are identified by measuring with 35 $\mu$m intrinsic resolution\footnote{The intrinsic resolution combined
with the beam-width $\sigma_{\rm{beam}}\simeq 30$ $\mu$m determines
 the total impact parameter resolution,  $\sigma_{\rm{SVT}}\oplus\sigma_{\rm{beam}}\simeq 47$ $\mu$m.} their impact parameter, which
is the minimum distance between  the particle direction and the primary \pap\ vertex in the plane transverse 
to the beam. Such a high accuracy can be reached only using online the silicon information, a challenging task that
requires to read-out 212,000 silicon channels and to complete hit-clustering and pattern recognition within the trigger latency.
In a highly-parallelized architecture, fast pattern-matching and
linearized track-fitting allow reconstruction of 2D-tracks in the plane transverse to the beam with offline-quality by combining
drift-chamber and silicon information, within a typical latency of 25 $\mu$s per event.\par
Using the above device, CDF implemented a trigger selection that requires only two displaced tracks in the event, 
to collect  pure samples of exclusive non-leptonic  $b$-decays for the first time in a hadron
 collider \eg\ \bdpipi and  $B^0_s\rightarrow D_s^-\pi^+$ decays, in addition to large samples of semileptonic $b$-decays 
and of \emph{charmed} meson decays.
 However, an impact-parameter based selection biases 
the decay-length distributions, reducing the statistical power in lifetime measurements. 
In section 3.2 we discuss how the original lifetime 
information is deconvolved from the trigger efficiency, and from the smearing effects 
due to the finite resolution on
the measured impact parameters and decay lengths.\par 
Besides the trigger on displaced tracks, past experience from Run I suggests that triggering on 
final states  containing  single or dileptons is a successful strategy
 to select samples of \bhadron\ decays, because semileptonic
 ($B\rightarrow l\nu_{l}X$)  and \emph{charmonium}  ($B\rightarrow \jpsi X\rightarrow [l^+l^-]X$) decays 
represent about 20\% of \bgmeson\ widths and have relatively clean experimental signatures.
Such a `conventional' approach was adapted to the upgraded detector: identification of muon down to low momenta allows for
 efficient dimuon triggers in which we select \emph{charmonium} or rare decays and then we fully reconstruct several decay modes.
  On the other hand, semileptonic triggers require a displaced track in addition to the muon (or electron), providing cleaner samples. 

\section{Analysis of \bhh\ decays}
The extraction of CKM parameters from measurements in \bhadron\ decays is often affected by large uncertainties, coming from 
non-perturbative QCD effects. One way to simplify the problem is to invoke flavor-symmetries under which 
the unknowns partially cancel. In this respect, joint study of \bd\ and \bs\ two-body decays into charged kaons and pions (\bhh) plays a
key role, since these modes are related by subgroups of the SU(3) symmetry \cite{bhh_1,bhh_2,bhh_3}.
 CDF is the only experiment, to date, that has simultaneous access to
these modes; this provides a rich flavor-physics program: 
 the analysis of first Run II data already lead to the first observation of \bskk\ decays, and to 
the world best limits on \bskpi\ and \bspipi\ decay rates \cite{bhh_4,bhh_5}; now, with  samples increased in size, CDF is approaching 
the opportunity to obtain competitive measurements of \CP-violating phases.
 
\subsection{\CP\ asymmetry in \bdkpi\ decays}
\label{ssec:acp}
The flavor-specific\footnote{We neglect the $\mathcal{O}(10^{-4})$ fraction of doubly-Cabibbo suppressed decays.}  \bdkpi\ decay 
occurs in the standard model (SM) through the dominant  `tree' (of amplitude T) and `penguin' (P) diagrams.  Their interfering amplitudes induce the
\CP\ asymmetry \acpbdkpi\ defined as follows:
\begin{eqnarray}
\frac{\br(\abdkpi) - \br(\bdkpi)}{\br(\abdkpi) + \br(\bdkpi)} \nonumber \\
 = \frac{2|T||P|\sin(\delta)\sin(\gamma)}{|T|^2 + |P|^2 + 2|T||P|\cos(\delta)\cos(\gamma)},
\end{eqnarray}
which is sensitive to the \vub\ CKM phase (angle $\gamma \equiv \phi_3$) and to the difference $\delta$ between strong phases of the two amplitudes. 
$B$-factories recently measured a $\mathcal{O}(10\%)$ asymmetry with 2\% accuracy, probing for the first time direct \CP\ violation 
in the \bquark\ sector \cite{acp_1,acp_2}; however, additional experimental information is needed because theoretical predictions still 
suffer from large (5--10\%) uncertainties \cite{acp_3,acp_4,acp_5}, and the observed asymmetries in neutral and charged modes are not consistent, 
as the SM would suggest. A measurement from the Tevatron is therefore interesting, also for the unique possibility to combine asymmetry measurements
in \bdkpi\ and \bskpi\ decays, which provide a model-independent probe for the presence of non-SM physics \cite{acp_6}.   
\par We analyzed a $\int\Lumi dt\simeq 360$ \lumipb\ sample of pairs of oppositely-charged particles, used to form \bnmeson\ candidates, 
 with $p_T > 2$ \pgev\ and   $p_T(1) + p_T(2) > 5.5$ \pgev. The trigger required also a transverse opening-angle between tracks
 $20^\circ < \Delta\phi < 135^\circ$ to reject background from particles within  light-quark jets.  
In addition, both charged particles were required to originate in 
a displaced vertex (100 $\mu$m $< d_0 < 1$ mm), while the \bnmeson\ candidate was required to be produced in the primary \pap\ interaction 
($d_0(B)< 140$ $\mu$m) and to have travelled a transverse distance $L_{xy}(B)>200$ $\mu$m.\par
\begin{figure}[h]
\centering
\includegraphics[width=40mm]{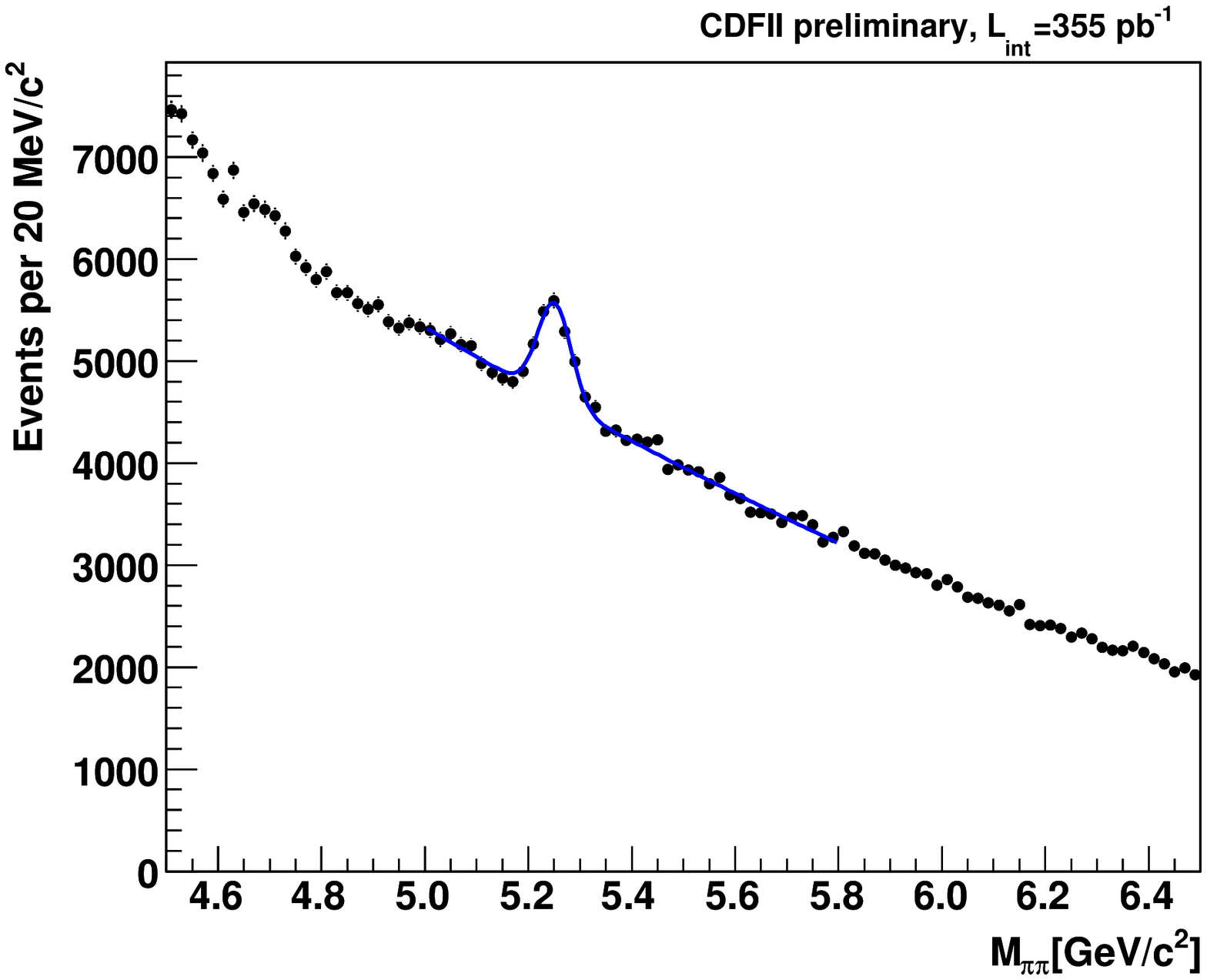}
\includegraphics[width=40mm]{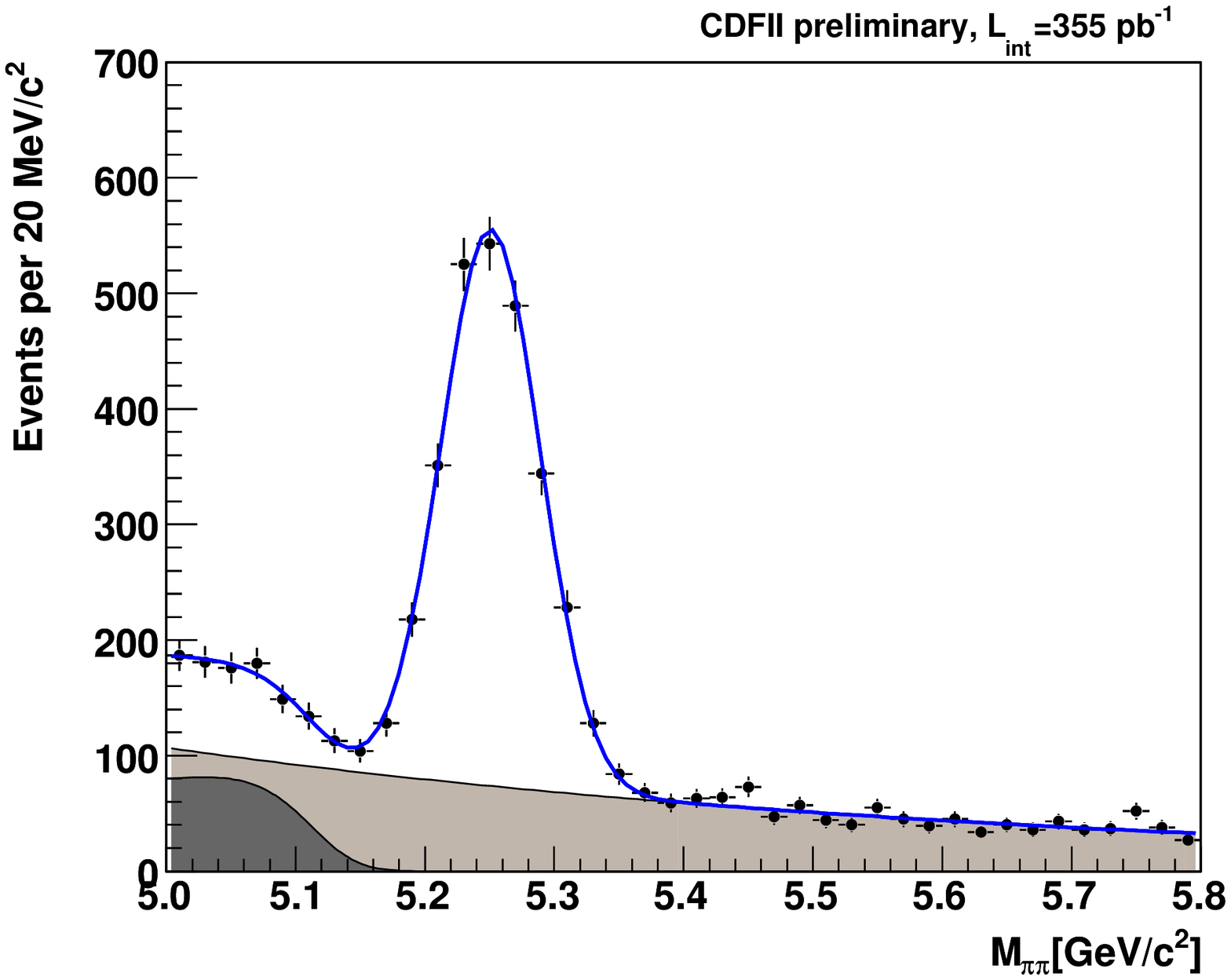}
\caption{Invariant $\pi\pi$-mass after the trigger (left plot) and after the optimized offline selection (right plot). 
Latter plot shows also background contributions from random pairs of tracks which satisfy the selection requirements (light gray) 
and partially-reconstructed \bnmeson\ decays (dark gray), as resulting from the invariant-mass fit.}
 \label{fig:acp_1}
\end{figure}

 A \bhh\ signal of about 3800 events and 
signal-to-noise ratio $\rm{SNR} \simeq 0.2$ at peak is visible in data (\fig{acp_1}, left)  already after confirming the trigger selection on offline quantities: extraction of a
$\br\simeq 10^{-5}$ signal at trigger-level is a remarkable achievement at a hadron collider, made it possible by the CDF trigger 
on displaced tracks.\par In the offline analysis, an unbiased procedure of optimization determined a
 tightened selection on track-pairs fit to a common decay-vertex. We found the optimal selection by minimizing the
 following analytical parameterization of the average expected resolution on the asymmetry measurement: 
$\sigma_{\acp} = \frac{1}{\sqrt{S}}\sqrt{z + w(B/S)}$.  For each set of cuts, $S$ was the signal yield, estimated from Monte Carlo 
simulation and normalized to the yield observed in data after the trigger selection, and $B$ were the background events found in the sidebands 
of the $\pi\pi$-mass distribution in data. The constants $w$ and $z$ were parameters determined \emph{a priori} from full analyses of pseudo-experiments
reproducing the experimental circumstance of data. By using $\sigma_{\acp}$ we obtained about 10\% improvement 
in resolution over the standard $\sqrt{S+B}/S$. Besides tightening the trigger cuts, in the analysis we exploited the discriminating power 
of the \bnmeson\ `isolation' and of the information provided by the 3D reconstruction capability  of CDF tracking, 
which both allowed great improvement in signal
 purity. Isolation was defined as $I(B)= \ptb/[\ptb + \sum_{i} \pt(i)]$, in which the sum runs over every other track within 
a cone of radius one   in the $\eta-\phi$ space around the \bnmeson\ flight-direction; by requiring $I(B)> 0.5$,  we reduced the background by a factor 
four while keeping almost 80\% of signal. The 3D view of tracking allowed to resolve multiple vertices along the beam direction 
reducing the background (mainly pairs of displaced tracks coming  from distinct, uncorrelated heavy-flavor decays) by a factor 
two, with little inefficiency on signal.
\par The resulting $\pi\pi$-mass distribution  (\fig{acp_1}, right) shows a clean signal, estimated by a Gaussian (signal)
plus an exponential (combinatoric background) and an Argus-shaped (physics background) fit to contain $2327 \pm 77$ events,
 with standard deviation $\sigma = 39 \pm 1$ MeV/$c^2$ and $\rm{SNR} \simeq 6.5$ 
at peak. This corresponds to a factor 1.7 reduction in signal yield and to a factor of 50 reduction in background with respect to the 
trigger selection (\fig{acp_1}, left).\par
Despite the excellent mass resolution, the various \bhh\ modes overlapped into an unresolved mass peak, while the PID resolution was
insufficient for separating them on an event-by-event basis. We achieved a statistical separation instead, with a
 multivariate, unbinned likelihood-fit (fit of composition) that used PID information, provided by the \dedx\ in the drift chamber, and kinematics.\par
\begin{figure}[h]
\centering
\includegraphics[width=40mm]{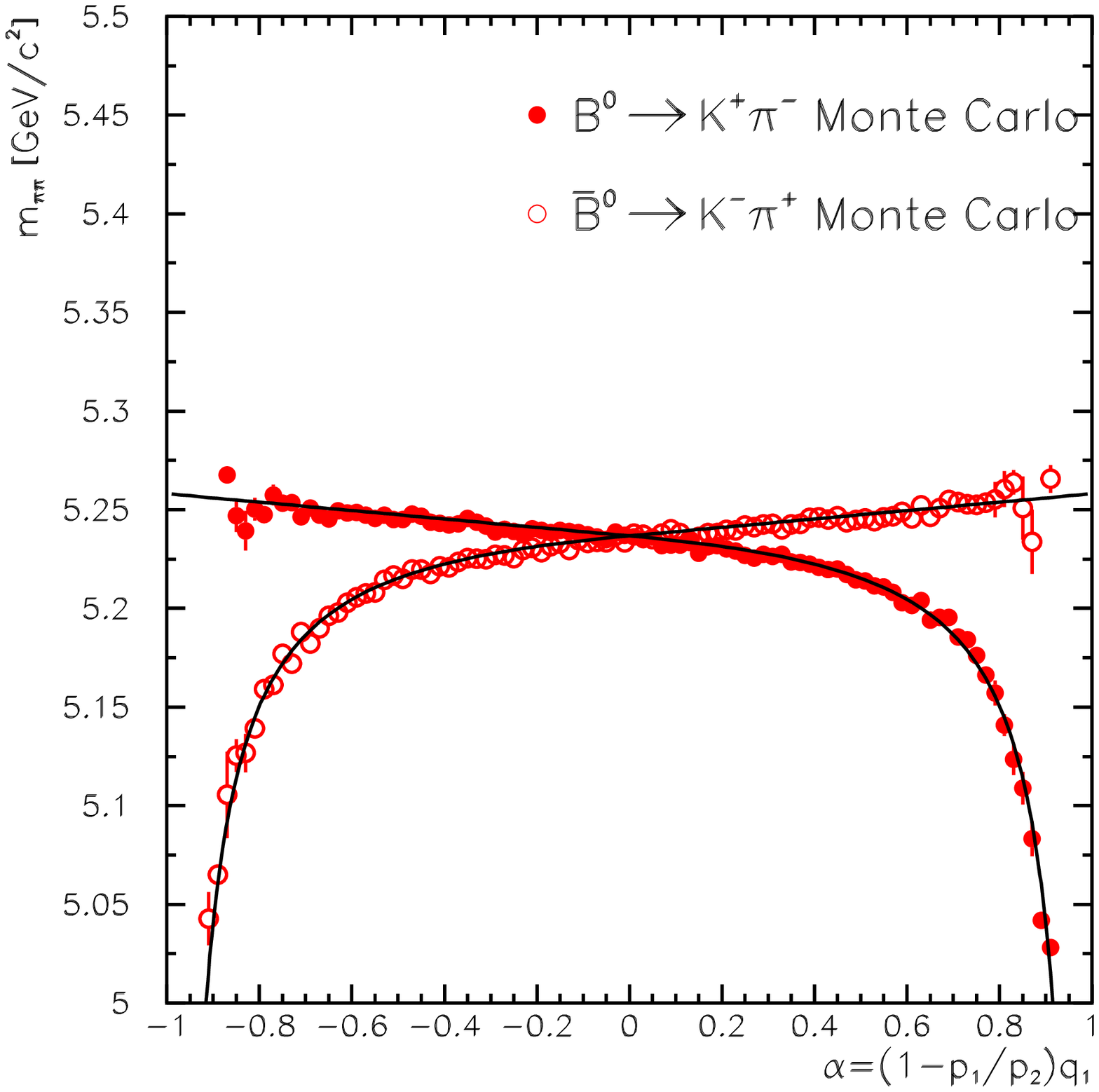}
\includegraphics[width=40mm]{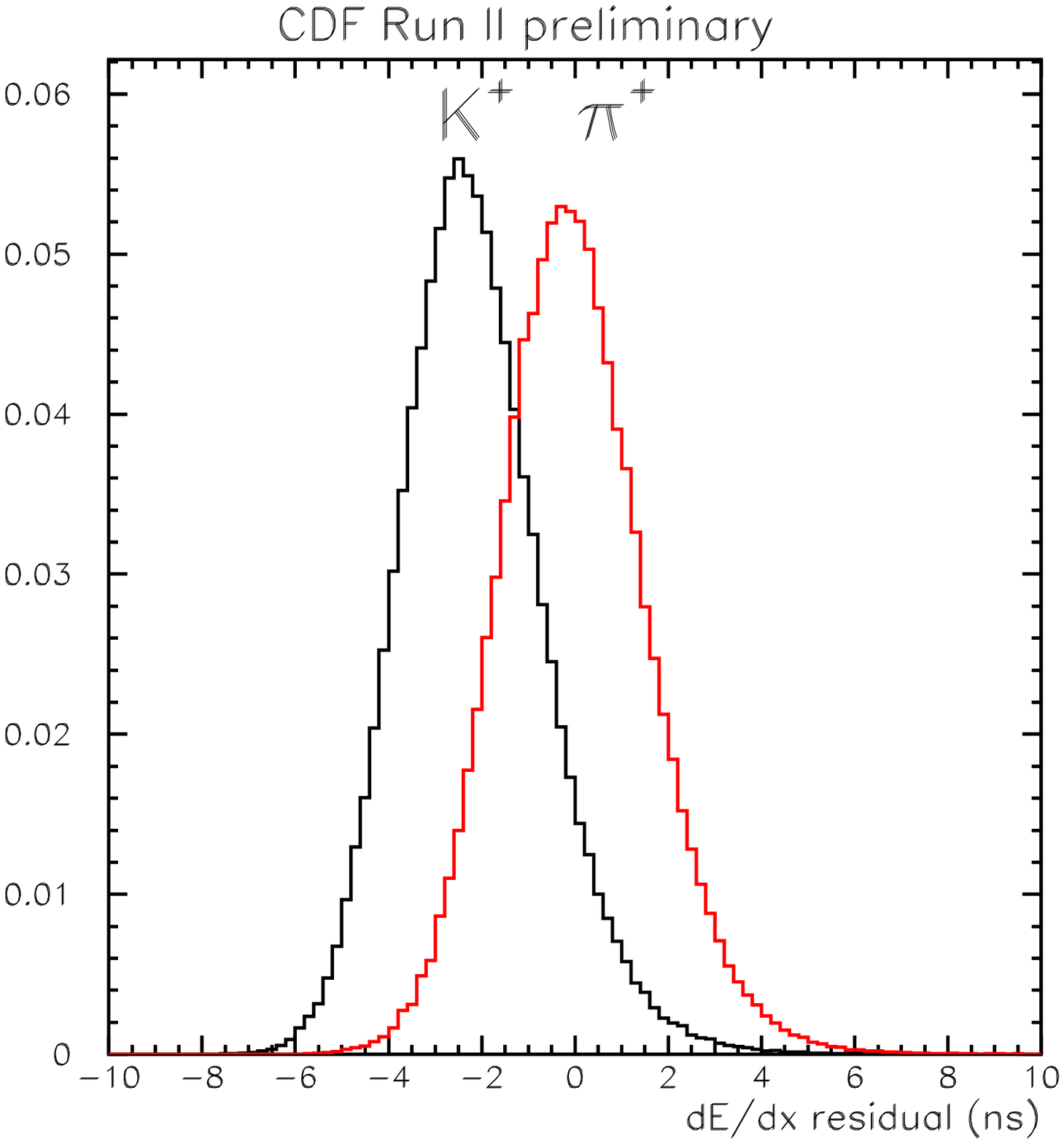}
\caption{Averaged $m_{\pi\pi}$ distribution as a function of the signed momentum imbalance ($\alpha$) for simulated \bdkpi\ and \abdkpi\ events with 
analytical function of \eq{acp_1} overlaid (left plot); distribution of the difference between observed \dedx\ and expected \dedx\ (in pion mass-hypothesis)
for positive kaons and pions (right plot).}
 \label{fig:acp_2}
\end{figure}

We exploited the kinematic differences among modes by using an approximate relation between any
 two invariant masses $(M_{m_{1},m_{2}}$ and $M_{m_{1}',m_{2}'})$ obtained with two 
arbitrary mass-assignments to the outgoing particles ($m_1, m_2$ and $m_1'$ $m_2'$).
 If $m_{1,2} \ll p_{1,2}$ we have
\begin{eqnarray} 
\label{eq:acp_1}
M^2_{m_1,m_2}\approx M^2_{m_1',m_2'}&+&\left(1+ p_1/p_2\right)\cdot\left(m_2^2 -m_2'^2\right) \nonumber \\ &+&\left(1+ p_2/p_1\right)\cdot\left(m_1^2 -m_1'^2\right),
\end{eqnarray}
where kinematic information associated to all possible mass-assignments ($K^+\pi^-$, $K^-\pi^+$, $\pi^+\pi^-$, $K^+K^-$) 
was compacted within just two observables,  a single candidate invariant-mass and the ratio of momenta.  Left plot in
 \fig{acp_2} shows  the averaged  $\pi\pi$-mass as  a function of  the signed momentum-imbalance, $\alpha=(1-p_1/p_2)q_1$,
 for simulated \bdkpi\  and \abdkpi\ events. The momentum (charge) $p_1$ ($q_1$) 
refers to the softer track; by combining kinematics and charge, we therefore separated also 
$K^+\pi^-$ from  $K^-\pi^+$ final states.\par We equalized the \dedx\ over the tracking volume and time using  
 about $760,000$  $D^{*+}\rightarrow D^0\pi^+\rightarrow [K^-\pi^+]\pi^+$ decays, 
where identity of \dzero\ daughters was tagged by the strong $D^{*+}$ 
decay \cite{acp_7}. The $\mathcal{O}(0.4\%)$  contamination from  doubly Cabibbo-suppressed $D^0\rightarrow K^+\pi^-$ decays was neglected.  
In a $>95\%$ pure \dzero\ sample, we obtained $1.4\sigma$ separation between kaons and pions (\fig{acp_2}, right), corresponding to an 
uncertainty on the measured fraction of each class of particles that is just 60\% worse than the uncertainty attainable with ideal 
separation.  We measured, and included in the fit, a  11\% residual track-to-track correlation due to 
common-mode \dedx\ fluctuations.\par
The fit of composition used five observables: 
the invariant $\pi\pi$-mass $m_{\pi\pi}$, the signed momentum-imbalance $\alpha$, the scalar sum 
of particles' momenta \ptot, and the \dedx\ of both particles. 
The likelihood for the single event $i$ was
$\like^i = (1-b)\sum_{j}f_j\like_j + b~\like_{\mathrm{bck}}$,
 where $j$ runs over the signal modes, and  $f_j$ ($b$) are the
 fractions of each mode (background) to be determined by the fit. 
In terms of probability density functions (p.d.f.), each term of the likelihood reads as
\begin{equation}
\label{eq:acp_2}
\like \simeq \pdf^{m}(m_{\pi\pi}|\alpha)\pdf^{p}(\alpha,\ptot)\pdf^{\rm{PID}}(\dedx_1,\dedx_2). 
\end{equation}
The mass ($\pdf^{m}$) model was obtained from the analytical formula \eq{acp_1} for signal and from mass-sidebands of data for background;
 the momentum term  ($\pdf^{p}$) was derived from simulation for signal and from mass-sidebands of data for background. The PID term ($\pdf^{\rm{PID}}$)
was extracted from calibration samples of \dzero\  decays for signal and background. 
\par
\begin{figure}[h]
\centering
\includegraphics[width=80mm]{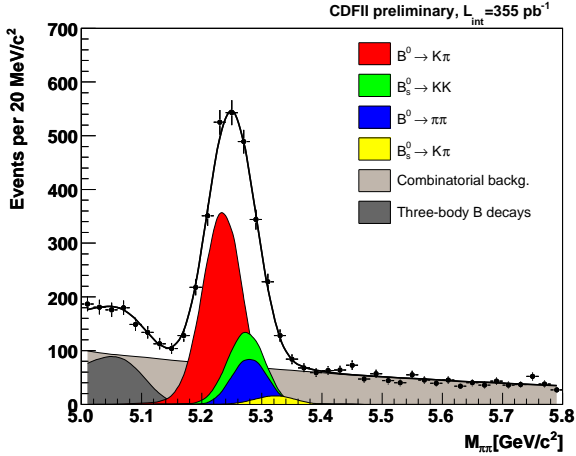}
\caption{Invariant $\pi\pi$-mass after the offline  selection with individual signal components (cumulative) and backgrounds (overlapping) 
overlaid.}  
 \label{fig:acp_3}
\end{figure}

 The fit found three modes that contribute to the peak:
$313 \pm 34$ \bdpipi, $1475 \pm 60$ \bdkpi,  and $523 \pm 41$ \bskk\ decays. A not yet statistically significant contribution 
of $64 \pm 30$ \bskpi\ decays was also found. Fit projections are overlaid to data in \fig{acp_3}.
\par From $787 \pm 42$ reconstructed \bdkpi\ decays and $689 \pm 41$ reconstructed \abdkpi\ decays, we measured the following uncorrected
 value for the direct \CP\ asymmetry:
\begin{equation}
\frac{N(\abdkpi) - N(\bdkpi)}{N(\abdkpi) + N(\bdkpi)} =  (-6.6 \pm 3.9)\%. \nonumber 
\end{equation}
Above result was then corrected for differences in trigger, reconstruction, and selection efficiencies between \bdkpi\ and \abdkpi\ modes;
since the known, small ($< 2\%$) detector charge-asymmetry canceled out almost completely in this measurement, the only effect
that mattered was the different probability of interaction with the beam-pipe and tracker material between positive and negative kaons.
This required a $(1.00 \pm 0.25)\%$ correction to the observed asymmetry. The value of the correction  was extracted from simulation
and checked in unbiased kaons from \dpluskpipi\ decays triggered on the pion pair. The corrected \CP\ asymmetry in \bdkpi\ decay-rates
is therefore $\acpbdkpi= (-5.8 \pm 3.9)\%$.\par
The various contributions to the systematic uncertainty, evaluated with pseudo-experiments, sum to a total uncertainty of 0.7\%, still
 smaller than the statistical uncertainty. The dominant source was the uncertainty  on the \dedx\ model for kaons, pions, and
 track-to-track correlation.  This effect is expected to partially reduce as the size of 
calibration samples of \dzero\ decays increases. The second important contribution derived from the statistical uncertainty 
on the nominal value of \bnmeson\ masses, which enters in the analytical expression \eq{acp_1}. Since we use the \bnmeson\ masses
 measured by CDF,  this uncertainty will reduce with the increasing statistic of fully-reconstructed \bnmeson\ decays. Other 
relevant contributions came from the uncertainty on the invariant-mass resolution for each \bhh\ mode, from possible 
shifts of the global mass scale with respect to nominal masses, from possible charge-asymmetries in background 
that could fake an asymmetry in \bdkpi\ rates, and from the uncertainty on the invariant-mass shape assumed for the combinatoric
 background.\par We quote the following result for the direct \CP\ asymmetry in \bdkpi\ decays, where all contributions to 
the systematic uncertainty have been summed in quadrature:
\begin{equation}
\acpbdkpi = (-5.8 \pm 3.9 \pm 0.7)\%,
\end{equation} 
which is approximately $1.5\sigma$ different from zero, and in agreement with world best results: 
$\acpbdkpi = (-11.3 \pm 2.2 \pm 0.8)\%$, from the Belle Collaboration \cite{acp_2}, and  $\acpbdkpi = (-13.3 \pm 3.0 \pm 0.9)\%$
from the Babar Collaboration \cite{acp_1}.\par
CDF result is still limited by the statistic uncertainty; however, its systematic uncertainty, at the same level of
$B$-factories, is  promising:  with significantly more data already collected,  we expect to reduce the statistical uncertainty 
 down to 2.5\%, which will make CDF result competitive with $B$-factories soon.

\subsection{Time-evolution of  \bskk\ decays}
\label{ssec:bskk}
The recent, first result on the \bs\ flavor-oscillation frequency by CDF \cite{bskk_1} is consistent with SM predictions.
The SM predicts the mass difference between \CP\ eigenstates\footnote{As usually allowed in the SM, we neglect the
 mixing phase $-2\beta_s$, thus mass and \CP\ eigenstates coincide.} $\Delta m_s$ to be related to the width difference $\Delta\Gamma_s$  by the relation 
$\Delta\Gamma_s/\Delta m_s \simeq 0.003$ \cite{bskk_2}; hence an observed $\Delta\Gamma_s$ inconsistent  with above ratio  
would hint at non-SM, \CP-violating, new phases. A  way to probe $\Delta\Gamma_s$ is to measure the lifetime of \CP-specific \bs\ decays, and to combine it with the 
lifetime in flavor-specific decays to extract  $\Delta\Gamma_s/\Gamma_s$.\par In the same $\int\Lumi dt\simeq 360$ 
\lumipb\ sample used for the asymmetry measurement, we measured the time-evolution of 
 untagged \bskk\ decays, which are expected to be 95\% \CP-even eigenstates.
A similar unbiased optimization procedure, aimed at improving the resolution on lifetime-measurements, yielded an offline-selection
 based on transverse-momentum, impact-parameter, and vertex-quality requirements. 
The resulting signal was similar to the one shown in right plot of \fig{acp_1}, and contained 
about 2200 \bhh\ decays, with 5.0 approximate peak SNR.\par
\begin{figure}[t]
\centering
\includegraphics[width=80mm]{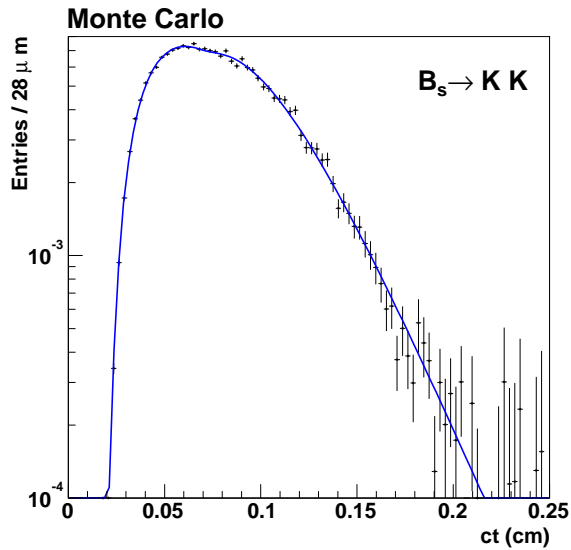}
\caption{Combined trigger and selection efficiency for the \bskk\ mode as a function of the decay length.} \label{fig:bskk_1}
\end{figure}

The time-evolution of individual signal modes  was determined by adding the decay-length information to the fit of composition 
described in the previous section. We multiplied the p.d.f. of \eq{acp_2} by the following lifetime-term:
\begin{equation}
\pdf^{\rm{life}}(ct) = [\mathrm{exp}(ct)\otimes\gauss(\sigma_{ct})]\times\epsilon(ct),
\end{equation}
in which the exponential modeled the evolution of the decay, the second term accounted for a Gaussian
detector-smearing that depended on the event-by-event uncertainty ($\sigma_{\mathit{ct}}$), and the third term
 was an efficiency that depended on the decay length and accounted for the lifetime bias.
The bias of the lifetime distribution due to resolution and efficiency effects introduced by the trigger on displaced vertices,
and by the offline analysis,  was modeled with a unique efficiency curve $\epsilon(ct)$.
This was defined as  the ratio between the pseudo-proper decay length distribution 
of events passing the trigger and the unsculpted one, and it was extracted from simulated samples.  We checked the agreement of simulation with
 real data, and we verified that the efficiency curves were independent
of the lifetime of the simulated samples used to derive them. A standard lifetime-fit was performed on
 $B^+\rightarrow \jpsi K^+$ decays, which were collected by the dimuon trigger, and therefore were free from any lifetime bias; 
 then we applied offline the selection of the displaced-track trigger to these decays to select a `lifetime-biased' sub-sample;
the lifetime of this sub-sample was fit using an efficiency curve extracted by the simulation, and  was found in good agreement with the
 unbiased lifetime of the whole sample. In addition, distinct samples of 
$B^+ \rightarrow \overline{D}^0 \pi^+$  were simulated with different lifetimes; an efficiency curve was extracted for each sample,
and was applied to a different lifetime-fit in data; no dependence on the $c\tau$ of the simulated sample
was found in the resulting lifetimes in data. The efficiency curve of the \bskk\ mode is shown in \fig{bskk_1}.
\par The addition of lifetime information in the fit did not biased the estimated
signal composition, which was found in good agreement with what obtained in the previous section. 
 The estimated pseudo-proper decay-lengths  were $c\tau(\bd) = 452 \pm 24$ $\mu$m for the \bdmeson, and $c\tau(\bskk) = 463 \pm 56$ $\mu$m for the \bskk\ decay. We assumed the time-evolution of
 the \bskk\ decay to be a single-exponential \ie\ \bskk\ was assumed to be a pure \CP-even eigenstate. \par
\begin{figure}[h]
\centering
\includegraphics[width=80mm]{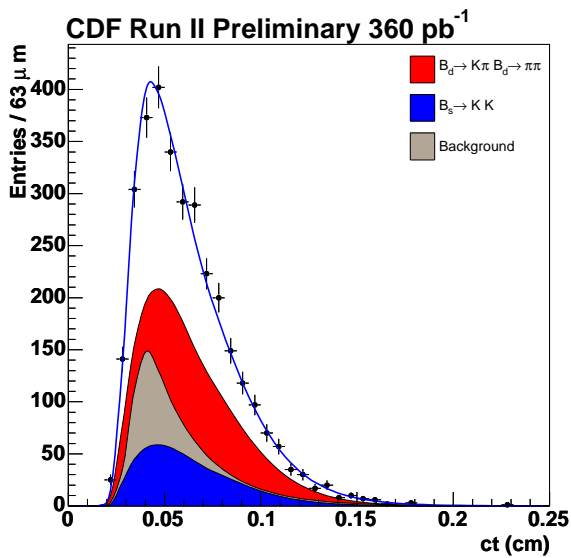}
\caption{Lifetime distribution of signal events (dots) with fit projections overlaid.} \label{fig:bskk_2}
\end{figure}

With a Gaussian constraint on the
\bdmeson\ lifetime to the world average value \cite{pdg}, we obtained $c\tau(\bskk) = 458 \pm 53$ $\mu$m. The lifetime distribution of the signal with fit projection overlaid is shown in \fig{bskk_2}.
\par The total systematic uncertainty was 5.6 $\mu$m, significantly smaller than the statistic uncertainty. Its dominant sources 
included the effect of misalignments in the tracker,  the uncertainties on the model used for proper-time resolution and for the lifetime of  background, the uncertainty on the \dedx\ model, 
the uncertainty on the input \ptb\ spectrum used in the simulation,  and the uncertainty on the extraction of the trigger-efficiency curve. 
The resulting \bskk\ lifetime was
\begin{equation}
\tau_L = 1.53 \pm 0.18 \pm 0.02~\rm{ps},  
\end{equation}
which, combined  with the world average flavor-specific \bsmeson\ lifetime \cita{hfag}, yielded the following
measurement of the width-difference in the \bs-system for \CP-eigenstates:
\begin{equation}
\frac{\Delta\Gamma^{\CP}_s}{\Gamma^{\CP}_s} = -0.08 \pm 0.23 \pm 0.03. 
\end{equation}
This result,  still limited by statistical uncertainty, is already the second world best measurement; an uncertainty at the 0.10 
level is expected with the upcoming upgrade of the analysis to the data already collected.
  
\section{Search for the rare \bmumu\ decays}
\label{sec:bmumu}
In the SM, Flavor Changing Neutral Current (FCNC) decays are strongly suppressed; for instance, the expected branching-fractions of rare
  \bmumu\ decays are around $3.4 \times 10^{-9}$ for the \bs\ mode \cite{bmumu_1,bmumu_2}, and approximately  $1 \times 10^{-10}$ \cita{bmumu_1} for the
 \bd\ mode, further suppressed by a factor $|\vtd/\vts|^2$. Above rates are a factor $\mathcal{O}(100)$ beyond current experimental sensitivity
 at the Tevatron.  However, contribution from non-SM physics may significantly enhance these rates, making it possible  an observation 
that would be unambiguous signature for new physics.\par In minimal supersymmetric (SUSY) extensions of the SM, for instance,
 additional processes involving 
virtual SUSY particles imply $\br(\bmumu)\propto \tan^6(\beta)$; $\tan(\beta)$ is the ratio of vacuum 
expectation  values of the two neutral \CP-even Higgs fields; hence large enhancements of the decay rate are expected
in those SUSY models, like minimal SO(10) \cite{bmumu_3,bmumu_4,bmumu_5,bmumu_6}, that favor higher values of $\tan(\beta)$ \cite{bmumu_7,bmumu_8,bmumu_9}.
 On the other hand, R-parity violating SUSY models \cite{bmumu_9}
may enhance \bmumu\ rates even at lower values of $\tan(\beta)$. Hence, while an observation of \bmumu\ decays would provide crucial information 
on the flavor-structure of new physics, even improved exclusion-limits constrain the available space of parameters of
 several SUSY models.\par
\begin{figure}[h]
\centering
\includegraphics[width=80mm]{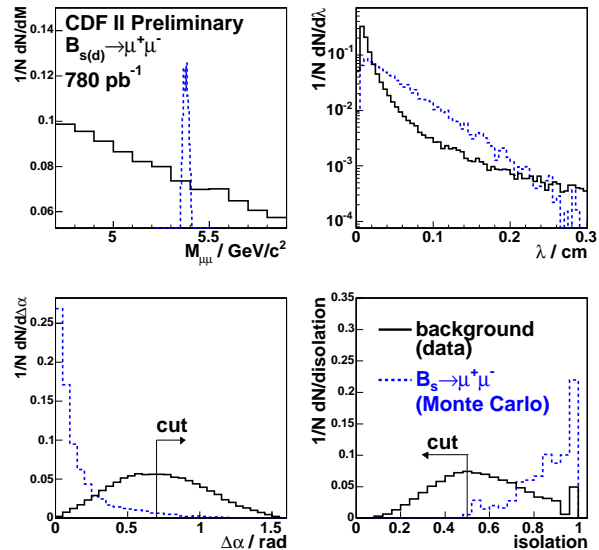}
\caption{Distribution of the discriminating observables for data, dominated by background (solid histogram), and simulated \bsmumu\ decays
 (dashed histogram).} \label{fig:bmumu_plots}
\end{figure}

 We searched for \bmumu\ decays in $\int\Lumi dt\simeq 780$ \lumipb\ of data collected by two dimuon triggers: one that required 
 both muon candidates in the $|\eta|<0.6$ region (U-U channel), the other that required one muon candidate 
  in the $0.6<|\eta|<1.0$ region (U-X channel). The offline selection required two oppositely-charged muon candidates fit to a common 
decay-vertex. 
We cut on the dimuon transverse momentum to reject combinatoric background, on the 3D decay-length ($\lambda$) and on its resolution 
 to reject prompt background, and on the isolation of the \bnmeson\ candidate to exploit the harder fragmentation of \bgmesons\ with respect to
 light-quark background;  in addition,  we required the candidate to point back to the primary vertex to further reduce combinatoric background and
 partially reconstructed \bhadron\ decays. The final sample contained about 23,000 candidates, mostly coming from random 
combinatoric background.  Distributions of the discriminating observables for signal (detailed simulation) and background 
(data) are shown in \fig{bmumu_plots}. 
\par
\begin{figure}[t]
\centering
\includegraphics[width=80mm]{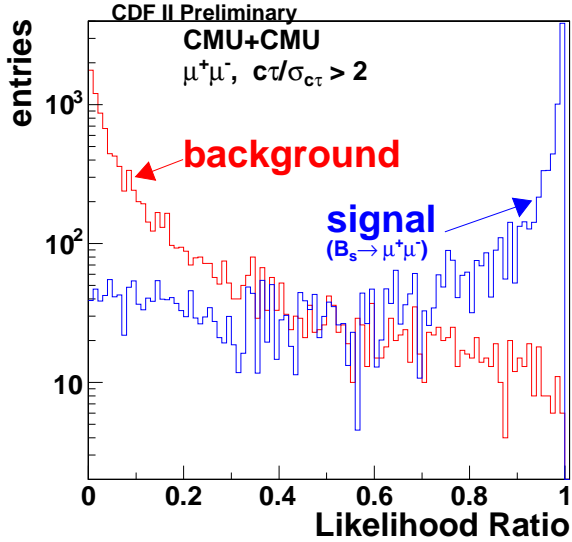}
\caption{Distribution of LR for data mass-sidebands (red histogram), and simulated signal (blue histogram).} \label{fig:bmumu_LR}
\end{figure}

 To further enhance purity we applied a cut on a likelihood-ratio (LR) variable based on three input observables: the isolation
 of the candidate, its decay-length  probability ($e^{-ct/c\tau}$), and its `pointing'  to the primary vertex
 (\ie\ the opening angle $\Delta\alpha$ between the \ptb-vector and the vector of 
 the displacement between the \pap\ vertex and the decay-vertex of the candidate). We extracted the signal p.d.f. from detailed simulation and the background p.d.f. from  mass-sidebands in data.
Distributions of LR for signal and background are shown in \fig{bmumu_LR}.\par
The \bmumu\ branching-fractions were obtained by normalizing to the number of $\bu \rightarrow \jpsi K^+ \rightarrow [\mu^+\mu^-]K^+$ 
decays collected  in the same sample. The ratio of trigger acceptances  
between signal and normalization modes ($\simeq 25\%$) was derived from simulation, the relative trigger efficiencies ($\simeq 1$) were extracted 
from unbiased data and the relative offline-selection efficiency ($\simeq 90\%$) was determined from simulation.\par
\begin{figure}[t]
\centering
\includegraphics[width=80mm]{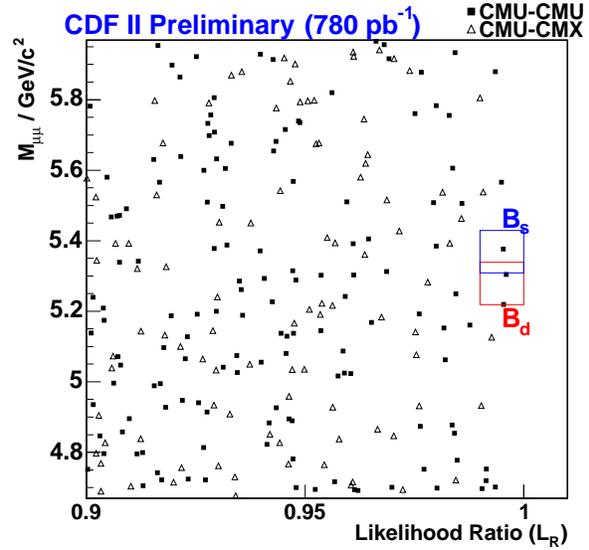}
\caption{Invariant $\mu^{+}\mu^{-}$-mass versus LR distribution for events satisfying the offline selection for U-U 
(solid square) and U-X (open triangle) channels. The \bs\ (blue box) and \bd\ (red box) signal  regions are also shown.} \label{fig:bmumu_box}
\end{figure}

The expected average number of background events was obtained by extrapolating events from the mass-sidebands to the search boxes.
 This estimate was checked  by comparing predicted and observed background yields in the following control samples: like-sign dimuon 
candidates, opposite-sign dimuon  candidates with negative decay-length, and opposite-sign dimuon candidates in which one 
muon failed the muon-quality requirements. Contributions of punch-through hadrons from \bhh\ decays were included in the estimate 
of total background. The optimal value for the LR cut was obtained by searching for the \emph{a priori} best expected 90\% C.L. 
upper limit on $\br(\bmumu)$. In two, 120 \massmev-wide search windows (to be compared with 25 \massmev\ mass-resolution) centered at
 the world average \bnmeson\ masses, we found 1 (0) and 2 (0) events in the U-U (U-X) channel for \bs\ and \bd\ decays 
respectively, in agreement with $0.88 \pm 0.30$ ($0.39 \pm 0.21$)  expected background events (see \fig{bmumu_box}).
The resulting combined upper-limits, estimated according to the Bayesian approach of \refcita{pdg}, and assuming a flat prior for the branching-fractions, 
were the following:
\begin{eqnarray}
\br(\bsmumu) < 8.0 \times 10^{-8}~\rm{at}~90\%~\rm{C.L.} \\
\br(\bdmumu) < 2.3 \times 10^{-8}~\rm{at}~90\%~\rm{C.L.}, 
\end{eqnarray}
which became respectively $\br(\bsmumu) < 1.0 \times 10^{-7}$ and $\br(\bdmumu) < 3.0 \times 10^{-7}$
at 95\% C.L. These results improve by a factor of two previous limits and significantly
reduce the allowed parameter space for a broad range of SUSY models.
\section{Summary and outlook}

We used $\int\Lumi dt\simeq 360$ \lumipb\ of data collected by the upgraded Collider Detector (\cdfii) at the Fermilab
Tevatron collider to measure the following \CP-violating asymmetry in \bdkpi\ decays: $\acpbdkpi = (-5.8 \pm 3.9 \pm 0.7)\%$
in agreement with the $B$-factories, and with comparable systematic uncertainties. In the same sample, the time-evolution
of the \bskk\ decay was measured for the first time, yielding to the following measurement of the relative width-difference: 
$\Delta\Gamma^{\CP}_s/\Gamma^{\CP}_s = -0.08 \pm 0.23 \pm 0.03$. With  $\int\Lumi dt\simeq 1$ \lumifb\  already collected,
 we expect about 2.5\% statistical uncertainty on the asymmetry, and about 0.10 uncertainty on the width-difference, which would place 
these CDF measurements among world best results. In this significantly larger sample, we also expect a precise measurement of the \bskk\ rate,
and either observation of \bskpi, \bspipi, $\Lambda^0_b \rightarrow \proton K^-$, $\Lambda^0_b \rightarrow \proton\pi^-$ decays, or setting world best 
limits on their rates. The $\int\Lumi dt\simeq 1$ \lumifb\ sample is also the starting point for the 
time-dependent analysis of flavor-tagged \bhh\ decays,  which may  allow CDF to measure \CP-violating 
asymmetries in \bskk\ and \bdpipi\ decays.\par Moreover, CDF quotes the following 90\% C.L. limits on the branching-fractions 
of FCNC, rare \bmumu\ decays: $\br(\bsmumu) < 8.0 \times 10^{-8}$ and 
$\br(\bdmumu) < 2.3 \times 10^{-8}$, obtained in the $\int\Lumi dt\simeq 780$ \lumipb\ sample. These are already the world best results, 
and they contributed to exclude a broad portion of parameter space in several SUSY models.

\bigskip 
\begin{acknowledgments}
I would like to thank the local organizers for a very enjoyable conference, the other participants for useful discussions, 
and the colleagues from CDF for helpful suggestion in preparing the talk and this document.
\end{acknowledgments}

\bigskip 

\end{document}